\documentclass[twoside]{article}

\usepackage[accepted]{aistats2019}

\usepackage[round]{natbib}

\usepackage{amsmath,amssymb}
\usepackage{graphicx}\usepackage{dcolumn}\usepackage{bm}\usepackage{amscd,amsthm}
\usepackage{xcolor}
\usepackage{ifthen}
\usepackage{balance}

\newcommand{\ve}[1]{\mathbf{#1} } 

\newcommand{\mc}[0]{\mathcal }

\newcommand\norm[2][\Tnorm]{\ensuremath{{\left\Vert #2 \right\Vert}_{#1}}}

\newcommand\ind[2][\Tind]{\ensuremath{ {\mathbf{1}_{#1} (#2) } }}

\newcommand\Tinnerprod{}
\newcommand{\innerprod}[3][\Tinnerprod]{\ifthenelse{\equal{#1}{}}{\ensuremath{\left<#2,#3\right>}}{\ensuremath{\left<#2,#3\right>_{#1}}}}

\newcommand\vect[1]{\mathbf #1}

\usepackage{hyperref}

\usepackage{tikz}
\usetikzlibrary{matrix}
\usetikzlibrary{positioning}
\usetikzlibrary{trees}

\usepackage[vlined,ruled,linesnumbered]{algorithm2e}

\newcommand\PR[2][\Tex]{
\ifthenelse{\equal{#1}{}}{{\mathrm{P}}\left[#2\right]}{\ensuremath{{\mathrm{P}}_{#1}\left[ #2\right]}}}

\newcommand\Tex{}
\newcommand\EX[2][\Tex]{
\ifthenelse{\equal{#1}{}}{{\mathbb E}\left[#2\right]}{\ensuremath{{\mathbb E}_{#1}\left[ #2\right]}}}

\newcommand\Var[2][\Tex]{
\ifthenelse{\equal{#1}{}}{{\mathrm{Var} }[#2]}{\ensuremath{\mathrm{Var}_{#1}\left[ #2\right]}}}

\newcommand\ignore[1]{}

\newcommand\alg{\mc A} 
\usepackage{mathtools}
\newcommand\defeq{\coloneqq}

\newcommand{\reals}{\mathbb R}

\newtheorem{theorem}{Theorem}
\newtheorem{lemma}[theorem]{Lemma}

\newtheorem{fact}[theorem]{Fact}

\newcommand\Ohidinglogs{\widetilde{O}}
\newcommand\N{N}

\renewcommand{\d}{d} 
\newcommand\eventscore{\mc E_\alpha} 
\newcommand\bad{\mc E_{\mathrm{bad}}}
\newcommand\timeind{t}

\newcommand\NCb{\widetilde \NC}
\newcommand\hatS{\widehat{\setS}}

\newcommand\setS{\mc S}
\newcommand\setX{\mc X}

\newcommand\mystackrel[2]{\stackrel{\text{#1}}{#2}}

 \renewcommand{\d}{d}  

\usepackage{bm}

\renewcommand{\ve}{\vect{e}}

\newcommand{\vu}{\vect{u}}

\newcommand{\vx}{\vect{x}}  
\newcommand{\vy}{\vect{y}}

\newcommand{\mQ}{\vect{Q}}

\newcommand{\mU}{\vect{U}}

\renewcommand{\S}{\setS}

\renewcommand\k{k}
\newcommand\h{h}
\newcommand\n{n}
\newcommand\close{\text{close}}
\renewcommand\middle{\text{middle}}
\newcommand\far{\text{far}}

\newcommand\dist{d}
\newcommand\hatdist{\widehat{d}}
\newcommand\NC{T}
\newcommand\dind{q}
\newcommand\bup{b}

\DeclareMathOperator*{\argmax}{arg\,max}
\DeclareMathOperator*{\argmin}{arg\,min}

\newcommand{\gap}[2]{\Delta_{#1,#2}}

\newcommand\nextind{\ell}

\begin{document}

\runningtitle{Adaptive Estimation for Approximate $k$-Nearest-Neighbor Computations}

\twocolumn[

\aistatstitle{Adaptive Estimation for \\ Approximate $k$-Nearest-Neighbor Computations}

\aistatsauthor{ Daniel LeJeune \And Richard G.\ Baraniuk \And Reinhard Heckel }

\aistatsaddress{ Rice University \And  Rice University \And Rice University } ]

\begin{abstract}
Algorithms often carry out equally many computations for ``easy'' and ``hard'' problem instances. In particular, algorithms for finding nearest neighbors typically have the same running time regardless of the particular problem instance.
In this paper, we consider the approximate $k$-nearest-neighbor problem, which is the problem of finding a subset of $O(k)$ points in a given set of points that contains the set of $k$ nearest neighbors of a given query point.
We propose an algorithm based on adaptively estimating the distances, and show that it is essentially optimal out of algorithms that are only allowed to adaptively estimate distances. 
We then demonstrate both theoretically and experimentally that the algorithm can  achieve significant speedups relative to the na\"ive method. 
\end{abstract}

\section{INTRODUCTION}
A large number of algorithms in machine learning and signal processing are based on distance computations. The algorithms for solving the associated computational problems are typically designed to perform well on a set of problem instances, in a worst-case or average-case sense, but do not necessarily have optimal or close-to-optimal computational complexity on any given problem instance. 
As a consequence, these algorithms often have running times and guarantees that are the same for ``easy'' and ``hard'' problems.

Ideally, we would like an algorithm that adapts to any given problem instance and only carries out the computations necessary for that problem instance. 
In this paper, we consider an approach for speeding up algorithms and evaluating their complexity that is based on adapting to a given problem instance with random adaptive sampling techniques~\citep{bagaria_medoids_2017,bagaria_kamath_tse_2018}. 
This approach is applicable to a variety of algorithms that are based on distance computations. Adding such adaptivity to algorithms can significantly speed up the algorithms' running times, since computationally easy problems have accordingly smaller running times. 

For concreteness, we focus on the problem of \emph{approximate} \emph{$k$-nearest-neighbor} ($k$-NN) computation.  Specifically, given a query point $\vx$ and a set of $n$ points, $\setX$, our goal is to find a subset containing the $k$ nearest neighbors of the query point. 
Our intuition is that an ``easy'' $k$-NN problem instance is one where there is a set of at least $k$ points that are close to the query point, and the other points are rather far, such that the evaluation of only a few coordinate-wise distances of the far points is sufficient to know that they are farther away than the close points. 
Contrarily, a ``hard'' problem instance is one where the distance from $\vx$ to all other points is very similar, and thus it is difficult to find a subset of $k$ nearest neighbors. 

We note that other formulations of the approximate $k$-NN problem are common as well. For example, \citet{Andoni_Indyk_2006} consider a formulation where points can be returned whose distance from the query point is within a multiplicative factor to its nearest points. 

We propose an algorithm, which we call the \emph{adaptive $k$-NN algorithm}, that adaptively estimates the distances and exploits the fact that for finding a set containing the $k$ nearest neighbors, it is not necessary to compute all distances exactly. In particular, for easy problem instances, coarse estimates are sufficient to identify a subset containing the $k$ nearest neighbors. 
Contrary to previous approaches, in particular that of~\cite{bagaria_kamath_tse_2018}, we focus on identifying a set \emph{containing} the $k$ nearest neighbors, since this is computationally considerably cheaper than identifying the \emph{exact} set of $k$ nearest neighbors.

We prove that the adaptive $k$-NN algorithm is \emph{near instance-optimal} for a restricted class of problems in the class of randomized algorithms that are based on adaptively estimating the distances.  
In a nutshell, the proof strategy is as follows: guaranteeing a solution to a given computational problem based on estimated distances (say, $k$-NNs) requires sufficiently good estimates of the distances. With standard change of measure techniques~\citep{kaufmann_complexity_2016} we can derive instance-dependent lower bounds on the sample complexity required to obtain sufficiently good estimates. 
This sample complexity is also a lower bound on the computational complexity of the respective algorithm, since at the very least the distances have to be computed sufficiently well to be sure a subset of the $k$ nearest neighbors can be identified with high probability. 
Further, we show that this computational complexity can also be achieved by designing an approximate $k$-NN algorithm that estimates the distances adaptively in time almost linear in the sample complexity. 
\section{RELATED WORK ON $k$-NN}

There are many highly efficient algorithms that solve versions of the (approximate) $k$-NN problem.
If the dimension of the data points is low, then the \mbox{k-d} tree algorithm~\citep{bentley_multidimensional_1975} performs very well. This algorithm builds a balanced k-d tree and traverses the tree to find the nearest neighbors. 
Contrary to our approach, the algorithm is based on pre-processing the data and thus becomes efficient only when performing many queries, so that the cost of building the tree becomes negligible. In addition, k-d trees become inefficient in high dimensions.

In order to overcome complexity in high dimensions, a number of works have proposed to find solutions that are approximate in that the algorithm is only asked to return points that are close in distance to the true nearest neighbors, for example, by using random projections~\citep{Ailon_Chazelle_2006}. 
Perhaps the most popular class of algorithms for performing approximate nearest-neighbor search is based on locality sensitive hashing~\citep{Andoni_Indyk_2006}. 
This class of algorithms works very well in theory and practice, but again uses a pre-processing step that is not negligible if only one query is executed.

If many $k$-NN queries are carried out on the same dataset, then the k-d algorithm for small dimensions and locality sensitive hashing algorithms for higher dimensions are significantly more efficient than algorithms based on adaptively estimating scores, such as the algorithm proposed here, since then the amortized pre-processing costs become negligible. In contrast, our approach is efficient in high dimensions and when we only carry out one or very few queries.

A setting particularly relevant to our approach is that in which the dataset is rapidly changing, where the assumption of other $k$-NN algorithms that pre-processing costs become negligible over time no longer holds. One such example is in the Impicit Maximum Likelihood Estimation procedure by~\citet{li2018implicit}, where at each iteration nearest-neighbor queries must be performed against a set of samples from the new estimate of the distribution.

There are a few recent success stories where adaptive randomized algorithms significantly speed up computational problems:
the Monte-Carlo tree search method for decision processes~\citep{kocsis_bandit_2006}, 
hyperparameter optimization in machine learning~\citep{li_hyperband_2016},
finding the medoid in a large collection of high-dimensional points~\citep{bagaria_medoids_2017}, and solving discrete optimization problems involving distance computations adaptively~\citep{bagaria_kamath_tse_2018}. 
All four works apply standard bandit algorithms in a creative way. 
Most related to our work is that of \cite{bagaria_kamath_tse_2018}, which proposes an efficient algorithms for solving the $k$-NN problem using an adaptive sampling strategy, similar to the one proposed here. The main difference is that we consider the approximate $k$-NN problem, which is a more general problem that contains the problem of finding the exact $k$-NN as a special case.
In order to solve the approximate $k$-NN problem, we have to solve a non-standard approximate bandit problem. 
In addition, we provide an algorithm that is \emph{near instance-optimal} in the class of algorithms that \emph{estimate} the distances for a restricted class of possible data points.

\section{PROBLEM STATEMENT}

Suppose we are given a set of high-dimensional points 
\[
\setX = \{\vx_1,\ldots,\vx_\n\} \subset \reals^m,
\]
and our goal is to find, for another given point $\vx \in \reals^m$, a set of size $O(\k)$ that includes the $\k$ nearest neighbors of $\vx$ to $\setX$ in $\ell_2$-distance (our results and discussion generalize to other distances, such as the $\ell_1$-distance).
This is a generalization of the exact $k$-nearest-neighbor problem and has applications in a vast number of classification tasks~\citep{hastie_elements_2009}. 

For convenience, we assume that all points are normalized such that $\|\vx\|_\infty \leq 1/2$, where $\norm[\infty]{\vx}$ denotes the largest absolute value of the vector $\vx$. 
We can brute-force solve the problem by computing all distances and then sorting, which yields a worst case complexity of $O(mn + n\log  n)$.
Our intuition is that it is unnecessary to compute the distances exactly, and that by approximating the distances we can save computations.

\section{THE ADAPTIVE $k$-NN ALGORITHM}

The idea behind our \emph{adaptive $k$-NN algorithm} is to adaptively estimate the distances. 
Then, the problem of finding a superset containing the $k$ nearest neighbors reduces to a \emph{multi-armed bandit} problem with the goal of identifying a set of size $k+h$ containing the $k$ smallest arms. 
Using that---up to a logarithmic factor---the sample complexity of the corresponding algorithm is equal to the computational complexity, we can provide an upper bound on the computational complexity of the algorithm by proving an upper bound on the sample complexity. 
Likewise, we can prove a lower bound for all algorithms that use adaptive estimates of the distances.

Our algorithm is inspired by the Hamming-LUCB algorithm in~\citep{heckel_approximate_2018} which in turn builds on the Lower-Upper Confidence Bound (LUCB) strategy, 
a popular algorithm for identifying the top-k items in a bandit problem~\citep{kalyanakrishnan_pac_2012} and for ranking from pairwise comparisons~\citep{heckel_active_2016}. 
The algorithm is based on actively identifying sets $\hatS_\close$ and $\hatS_\far$ consisting of $\k$ and $\n - \k - \h$ points, respectively, such that with high confidence the points in the first set have a smaller distance to $\vx$ than the points in the second set.
Once we have found such sets, the set
\[
\hatS = \{1,\ldots, \n \} \setminus \hatS_\far
\]
contains the closest $k$ points. Note that the cardinality of $\hatS$ is $\k+\h$, so to obtain a set of cardinality $O(k)$ conaining the $k$ nearest neighbors, we choose $\h$ on the order of $\k$. 
We adaptively estimate the normalized squared distances 
\[
\dist_i = \frac{1}{m} \norm[2]{\vx - \vx_i}^2
\]
by sampling indices $j_1,\ldots, j_{\NC_i}$ uniformly at random\footnote{Alternatively, one could select $j_1, \ldots, j_m$ uniformly at random without replacement, in which case $\hatdist_i(m) = \dist_i$ exactly. This could be implemented efficiently using ciphers, such as those described by \cite{black2002ciphers}.} from all indices $\{1,\ldots,m\}$ and then estimating the squared distance as
\[
\hatdist_i(\NC_i) = \frac{1}{\NC_i} \sum_{ j \in \{ j_1,\ldots, j_{\NC_i} \} }
| [\vx_{i}]_j - [\vx]_j|^2,
\]
where $[\vx]_j$ denotes the $j$-th coordinate of $\vx$.
The key idea is to estimate the distances only sufficiently well as to be able to obtain the two sets $\hatS_\close$ and $\hatS_\far$ from them.

Let $\NC_i$ be the counter of the number of samples used for estimating the respective distance. 
We define a confidence bound based on a non-asymptotic version of the
law of the iterated
logarithm~\citep{kaufmann_complexity_2016,jamieson_lil_2014}; it is of
the form\footnote{The constants involved can explicitly be chosen as 
$
 \alpha(u) = \sqrt{ \frac{
    2\beta(u,\delta/\n) }{ u} }\quad \text{with }
\beta(u,\delta') = \log(1/\delta') + 3\log\log(1/\delta') +
1.5\log(1+\log(u)).
$}
\[
\alpha(u) \propto \sqrt{ \frac{ \log( \log(u) \n/\delta) }{ u } },
\] 
where $u$ is an integer
corresponding to the number of samples, and $\delta$ is a parameter such that the algorithm succeeds with probability at least $1 - \delta$. 
Within each round, we also let $(\cdot)$ denote a
permutation of $[\n]$ such that \mbox{$\hatdist_{(1)} \leq
  \hatdist_{(2)} \leq \dots \leq \hatdist_{(\n)}$.} We then
define the indices
\begin{align}
\dind_1 &= \argmax_{i \in \{ (1), \ldots, (\k) \}} \hatdist_i + \alpha_i, \\
\dind_2 &= \argmin_{i \in \{ (\k+1+\h), \ldots, (\n) \}} \hatdist_i - \alpha_i,
\label{eq:d_ind}
\end{align}
where $\alpha_i = \alpha(T_i)$.
These indices are the analogues of the standard indices of the Lower-Upper Confidence Bound (LUCB) strategy from the bandit literature~\citep{kalyanakrishnan_pac_2012} for the bottom $\k$ and top $\n - \h - \k$ arms. 
The LUCB strategy for exact bottom-$\k$ recovery would update the scores $\dind_1$ and $\dind_2$ (for $\h=0$) at each round. Our strategy will go after what it ``thinks'' are the bottom $\k$ items, $\hatS_\close = \{(1), \ldots, (\k)\}$, and what it ``thinks'' are the top $\n -\k-\h$ items, $\hatS_\far = \{(\k+1+\h), \ldots, (\n) \}$.
Moreover, the algorithm keeps all the other items, $\hatS_\middle = \{(\k+1), \ldots, (\k+\h)\}$, in consideration for inclusion in these sets by keeping their confidence intervals below the confidence intervals of the items in $\hatS_\far$ (see \eqref{eq:blowbupdef} in the algorithm below). 
This is crucial to ensure that the algorithm does not get stuck trying to distinguish the middle items, which in general requires many samples.

\begin{algorithm}[ht]
\textbf{Input:} Confidence parameter $\delta$, approximation parameter $\h$ \\ \textbf{Initialization}: 
For every $i \in [\n]$, sample an index $j$ uniformly at random from $[m]$ and set 
$\hatdist_i(1) = | [\vx_{i}]_j - [\vx]_j|^2$, $\NC_{i} = 1$. \\
\textbf{Do} until termination:\\
\Indp Let $(\cdot)$ denote a permutation of $[\n]$ such that $\hatdist_{(1)} \leq \hatdist_{(2)} \leq \dots \hatdist_{(\n)}$. \\
For $\dind_1$ and $\dind_2$ defined by equation~\eqref{eq:d_ind}, define the index
\begin{align}
\label{eq:blowbupdef}
\bup_2 = \argmax_{i \in \{\dind_2,(k+1),\dots,(k+h)\}} ~ \alpha_i. 
\end{align}
\\
\textbf{For} $\nextind \in \{\dind_1, \bup_2\}$, increment $\NC_\nextind\leftarrow \NC_{\nextind}+1$, 
sample an index $j$ uniformly at random from $[m]$, 
and update 
$\hatdist_{\nextind} \leftarrow \frac{\NC_\nextind-1}{\NC_\nextind} \hatdist_\nextind + 
\frac{1}{\NC_\nextind} | [\vx_{\ell}]_j - [\vx]_j|^2$. 
If $T_\ell = m$, then compute the distance $\dist_\ell$ exactly and set $\hatdist_\ell = \dist_\ell$ and $\alpha_\ell=0$.
\\
 \textbf{End Loop} once the termination condition holds:
\begin{align}
\label{eq:termination}
\hatdist_{\dind_1} +\alpha_{\dind_1}
\leq 
\hatdist_{\dind_2} - \alpha_{\dind_2}.
\end{align}\\
\Indm
\textbf{Return} 
$\hatS = \{(1), \ldots, (k+\h) \}$ as an estimate of the set containing the $k$ nearest neighbors.
\caption{
\label{alg:HLUCB}
Adaptive $k$-NN}
\end{algorithm}

\subsection{Logarithmic Computational Complexity for Each Iteration}
\label{sec:iteration-complexity}

We next describe several implementation details that are critical for ensuring that each iteration of the adaptive $k$-NN algorithm has computational complexity $O(\log(n))$ and not $O(n \log(n))$, which a na\"ive implementation of computing the permutations via sorting and max and min computations would have. The key to achieving a logarithmic computational complexity is realizing that since only two distance estimates are updated in each iteration, the orderings of the distance estimates and confidence bounds will not change much between iterations, and at each iteration we only update the distance estimate for some indices that minimize or maximize some quantities relating to the confidence bound. This makes the algorithm amenable to the use of a \emph{heap data structure}~\citep{cormen2009introduction} to reduce computational complexity.

Figure~\ref{fig:heaps} illustrates how a total of seven heaps can be used to implement the adaptive $k$-NN algorithm efficiently. For each of $\hatS_\close$, $\hatS_\middle$, and $\hatS_\far$, a set of two or three coupled heaps is maintained, providing ordering information on both the distance estimates (so that we can maintain our permutation at each iteration) and the confidence bounds (so that we can select which distance estimates to update). For example, to determine $\dind_2$, we can simply extract the minimum from the min-heap on $\hatS_\far$ defined on the lower confidence bounds $\hatdist_i - \alpha_i$, which has a computational cost of $O(1)$. Later in the iteration, if we update the distance estimate at $\dind_2$, we update both the distance estimate min-heap and lower confidence bound min-heap on $\hatS_\far$ accordingly, which has a computational cost of $O(\log(n))$. At the end of the iteration, after making such updates across all three sets of heaps, the distance estimate ordering may not be maintained; e.g., the largest distance estimate in $\hatS_\middle$ may be larger than the smallest distance estimate in $\hatS_\far$. Items from each set must be swapped with items from other sets accordingly to restore the distance estimate ordering. Only two distance estimates are updated at each iteration, so at most a constant number of swaps that does not depend on $n$ are required, and each swap involves updating the appropriate heaps, yielding that the computational cost of restoring the distance estimate ordering is also $O(\log(n))$. Thus, the overall complexity per iteration is $O(\log(n))$. We ask the reader to refer to our implementation\footnote{See \url{https://github.com/dlej/adaptive-knn}.} for further details.

\newcommand\heap[1]{
\begin{tikzpicture}[scale=0.2,rotate=#1,transform shape,solid,level/.style={anchor=center,sibling distance=70pt/##1},circlenode/.style={solid,text width=1pt,circle,draw,minimum size=20pt}]
\node [circlenode,fill=black] (head) {}
    child {node [circlenode] (n10) {}
        child {node [circlenode] (n100) {}
            child {node [circlenode] (n1000) {}
            }
            child {node [circlenode] (n1001) {}
            }
        }
        child {node [circlenode] (n101) {}
            child {node [circlenode,draw=none] (n101x) {} edge from parent[draw=none]
            }
            child {node [circlenode] (n1010) {}
            }
        }
    }
    child {node [circlenode] (n11) {}
        child {node [circlenode] (n110) {}
        }
        child {node [circlenode] (n111) {}
        }
    };
\end{tikzpicture}
}

\newcommand\minheap{\heap{180}}
\newcommand\maxheap{\heap{0}}

\newcommand\coltwowidth{65pt}
\newcommand\colthreewidth{40pt}
\newcommand\rowheight{30pt}
\newcommand\sepwidth{2pt}
\newcommand\rightofbrace[1]{\rotatebox{-90}{$\overbrace{\hspace{\rowheight}}^{\rotatebox{90}{#1}}$}}

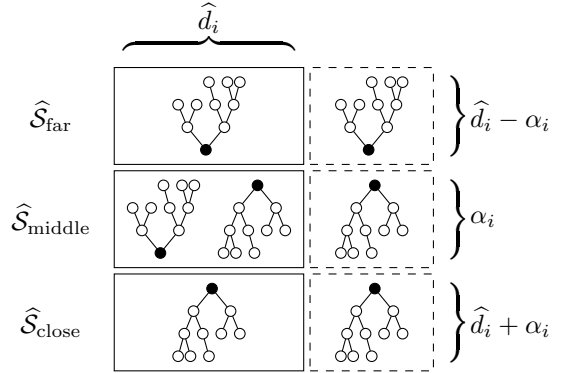
\begin{figure}[t]
    \centering
    \begin{tikzpicture}
    \matrix[
        matrix of nodes,align=center,
        row sep=\sepwidth,column sep=\sepwidth,
        nodes={anchor=center,align=center},
        column 2/.style={nodes={draw,text width=\coltwowidth}},
        column 3/.style={nodes={draw,dashed,text width=\colthreewidth}},
        column 4/.style={nodes={anchor=west,align=left}}
    ] (mat){
        & |[draw=none]| $\overbrace{\hspace{\coltwowidth}}^{\textstyle \hatdist_i}$ & 
        |[draw=none]| \\
        $\hatS_\far$ & \minheap & \minheap & \rightofbrace{$\textstyle \hatdist_i - \alpha_i$}  \\
        $\hatS_\middle$ & \minheap \maxheap & \maxheap & \rightofbrace{$\textstyle \alpha_i$} \\
        $\hatS_\close$ & \maxheap & \maxheap & \rightofbrace{$\textstyle \hatdist_i + \alpha_i$} \\
    };
    \end{tikzpicture}
    \caption{Illustration of how seven heaps can be used in the adaptive $k$-NN algorithm. Upward-branching trees indicate min-heaps and downward-branching trees indicate max-heaps.}
    \label{fig:heaps}
\end{figure}

\subsection{Guarantees and Optimality of the Adaptive $k$-NN Algorithm}

We next establish guarantees on the adaptive $k$-NN algorithm's success as well as on its computational complexity. 
The computational complexity depends on the gaps between the distances, defined as
$
\gap{i}{j} = \dist_i- \dist_j
$
through the function
\begin{align}
\label{eq:sampcsimple}
&\N(\vx,\setX,\h)
=
\Ohidinglogs \bigg(
\sum_{i = 1}^{\k}
\min(\gap{i}{\k+1+\h}^{-2},m) \\
&+
\sum_{i = \k+1+\h}^{\n} 
\min(\gap{\k}{i}^{-2},m) +
\h \min(\gap{\k}{\k+1+\h}^{-2},m)
\bigg). \nonumber
\end{align}
The notation $\Ohidinglogs$ absorbs factors logarithmic in $\n$ and doubly logarithmic in the gaps. 

\begin{theorem}
\label{thm:$k$-NN}
For any points $\vx$, $\setX$, the adaptive $k$-NN algorithm run with parameters $\delta$ and $\h$ yields a set of size $\k+\h$ that contains the $\k$ nearest neighbors and
has computational complexity at most 
\[
\N(\vx,\setX,\h)
\]
with probability at least $1 - \delta$.
\end{theorem}

Note that the computational complexity of the adaptive $k$-NN algorithm is upper-bounded by the complexity of the na\"ive brute force method, $O(mn + n\log  n)$, but is potentially significantly smaller. 
In particular, the computational complexity is small if there is a large gap between the $k$-th closest point and the \mbox{$(\k+\h)$-th} closest point. Taking $\h = \k$, for example, the algorithms returns a set of size $O(k)$ containing the $k$ nearest neighbors.
We next present two examples, one where the sample complexity of the adaptive $k$-NN algorithm is small, and one where it does not realize savings over the brute force method. 

Both of these examples assume the data lies in a low-dimensional linear subspace, a regime where one might consider using projection-based $k$-NN methods. However, such methods require knowledge of the dimension of the subspace and often have projection cost at least $O(mn)$. Even if the dimension is known and the projection is as efficient as possible (such as subsampling), our method will still have the advantage in that it adapts to the distances.

First, consider a $p$-dimensional subspace spanned by a matrix $\mU \in \reals^{m\times p}$ that has orthogonal columns. 
In addition, suppose that the columns of $\mU$ are incoherent with respect to the standard basis vectors $\ve_i$. Specifically, suppose that the maximum inner product between a column of $\mU$ and a standard basis vector obeys 
\[
\max_{i,j} \left| \innerprod{ \frac{\vu_i}{ \norm[2]{\vu_i} } }{\ve_j} \right| 
\leq   \frac{B}{\sqrt{m}}
\]
for some constant $B \geq 1$. We say $\mU$ is \emph{incoherent} if this bound holds for small values of $B$. 
Suppose then that the columns of $\mU$ are normalized to $\ell_2$-norm $\sqrt{m/p}/B$ and that the dataset $\setX$ and query point $\vx$ lie in that subspace, i.e., 
\begin{gather*}
\vx = \mU \vy,
\quad
\setX = \left\{ \mU \vy_i : \vy_i \in \mc{Y}  \right\},
\end{gather*}
where $\vy \in \reals^p$ and $\mc{Y} \subset \reals^p$ are the associated coefficient vectors. Assume that the associated coefficient vectors are normalized to have $\ell_2$-norm equal to $1/2$. 
Denote the gaps in the coefficient space by $\Delta_{i,j}^\vy$. 
From these assumptions, we are guaranteed that $\vx$ and the points in $\setX$ are bounded in $\ell_\infty$ norm by $1/2$. 
In addition, we have that
\begin{align*}
    \frac{1}{m} \norm[2]{\vx - \vx_i}^2 = \frac{1}{p B^2} \norm[2]{\vy - \vy_i}^2.
\end{align*}
Then $(\Delta_{i,j})^{-2} = (\Delta_{i,j}^\vy)^{-2} B^4$, so the computational complexity of the adaptive $k$-NN algorithm behaves like $\Ohidinglogs \left(n (\Delta_{k, \k+1+\h}^\vy)^{-2} B^4 \right)$ for large $m$; i.e., the computational complexity does not scale linearly with $m$. 
Hence, we can expect the adaptive $k$-NN algorithm to achieve significant computational savings when the data lies in a low-dimensional subspace of $\reals^m$. Furthermore, the algorithm is able to realize these savings \emph{without having this subspace or its dimension specified}. We illustrate this ability in our experiments below.

Next, suppose that the subspace is coherent with respect to the identitiy matrix. For example, consider the extreme case where all points lie in the one-dimensional subspace spanned by a single standard basis vector $\ve_i$. Then, estimation of the distances to an accuracy of $O(1)$ requires at least $O(m)$ samples, and thus the adaptive $k$-NN algorithm will always have the same sample complexity as the na\"ive brute force algorithm.

We next show that the algorithm is optimal among active algorithms that estimate the distances by sampling indices when the data points satisfy $[\vx]_j \in \{-1/2, 1/2 \}$. We note that it is only the coordinates of the data that are so constrained; the normalized distances themselves can be essentially any values between 0 and 1 for large enough $m$.

\begin{theorem}
\label{thm:necessity}
For any $\delta \in (0, 0.14]$, let $\alg$ denote an algorithm that can only interact with the data by sampling coordinates uniformly at random and yields, for any $\vx$ and $\setX$, the $\k$ nearest neighbors with probability at least $1-\delta$.
Then, when $\alg$ is run on any set of data points $\vx$, $\setX$ such that each coordinate of each point is either $-\frac{1}{2}$ or $\frac{1}{2}$, it has expected sample complexity at least 
\begin{align*}
\N_{\text{low}}(\vx,\setX,\h)
=
c'\left(
\sum_{i = 1}^{\k-\h} \gap{i}{\k+1+\h}^{-2} 
+
\sum_{i = \k+1+\h}^{\n} \gap{\k-\h}{i}^{-2}\right),
\end{align*}
where $c' = \log \left( \frac{1}{2\delta} \right) \min_{\ell \in \{\k -\h, \k + 1 + \h\}} \{ {\dist_{\ell} (1 - \dist_{\ell})} \} $.
\end{theorem}

Note that the above lower bound does not depend on the gaps involving the items $\k-\h+1,\ldots, \k+\h$. However, in the case when $\dist_{\k-\h} = \dist_\k$, we can relate the lower bound and the upper bound by 
\[
\N(\vx,\setX,2\h)
\leq \Ohidinglogs(
\N_{\text{low}}(\vx,\setX,\h)),
\]
so that we see that, up to rescaling of the approximation parameter $\h$ and logarithmic factors, the upper and lower bounds match. 
We emphasize that the lower bound only applies to algorithms that interact with the data by \emph{uniformly sampling} the distances and only when we constrain the data points. Thus, Theorem \ref{thm:necessity} only tells us that the adaptive $k$-NN algorithm is optimal among algorithms based on adaptively estimating the distances, but it does not state that algorithms based on other strategies cannot perform better. 

\section{EXPERIMENTS}

We run the adaptive $k$-NN algorithm both on artificial data restricted to low-dimensional subspaces and on real data to demonstrate the effectiveness in reducing computation over the na\"ive algorithm. The artificial data is generated via $\vx = c \mQ \vy$, where $\mQ \in \reals^{m \times p}$ is an i.i.d.\ Gaussian matrix, normalized to have unit-norm columns, $\vy$ are drawn uniformly at random from the unit sphere, and $c$ is the largest scalar such that $\norm[\infty]{\vx} \leq 1/2$ for all $\vx$ in the generated dataset. The real data comes from the Tiny ImageNet dataset \citeyearpar{tinyimagenet2015}, taking pixel values in $[0, 1]$. For each trial, we select $\setX$ by sampling $\n$ points at random (without replacement for Tiny ImageNet) and then select the query point $\vx$ by drawing another sample. For these experiments we used $\n = 1000$ and $m = 12288$, where the choice for $m$ comes from the dimensionality of the Tiny \mbox{ImageNet} images, which are $64 \times 64 \times 3$. Futher, we use $\alpha(u) = \sqrt{\frac{C_\alpha \log\left( 1 + (1 + \log(u)) n /\delta \right)}{u}}$ and vary $C_\alpha$.

\begin{figure}[t]
  \centering
  \includegraphics[width=\linewidth]{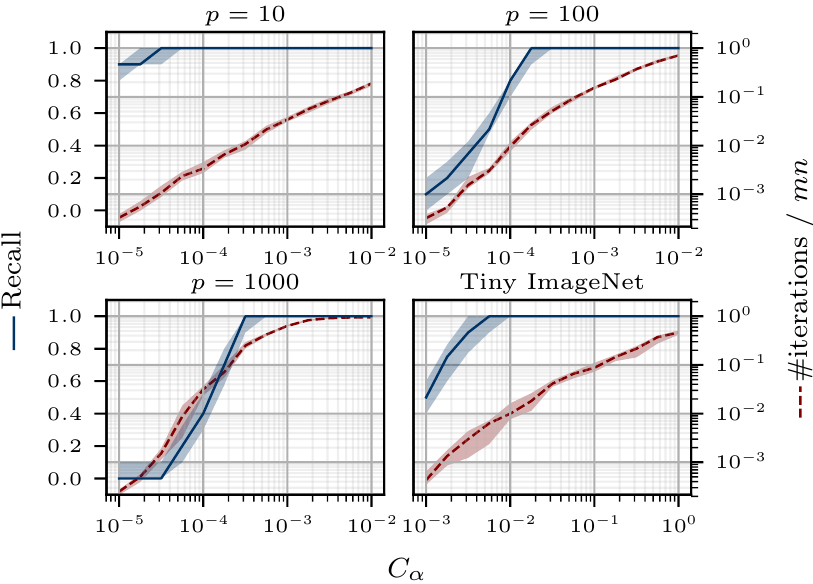}
  \caption{Effect of varying $C_\alpha$ for the adaptive $k$-NN algorithm on the random data lying in 10-dimensional (upper left), 100-dimensional (upper right), and 1000-dimensional (lower left) subspaces and on images from Tiny \mbox{ImageNet} (lower right). In each plot both the recall (blue, solid) and fraction of iterations used (red, dashed) are shown.}
  \label{fig:tiny_imagenet}
\end{figure}

Figure \ref{fig:tiny_imagenet} depicts the fraction of $\S_\close$ contained in $\hatS$ (recall) along with ratio of the number of iterations taken by the adaptive $k$-NN algorithm versus the na\"ive algorithm as we vary $C_\alpha$. Here $\k = 10$, $\h = 10$, and $\delta = 0.001$. We perform 20 random trials at each $C_\alpha$ and plot the median value (lines) and interquartile range (shaded area) over the trials. As we expect from our previous discussion, for low-dimensional subspaces (e.g., $p=10$) we see significant computational savings (multiple orders of magnitude) by using the adaptive $k$-NN algorithm over the na\"ive method while still being able to return a set containing all of the $\k$ nearest neighbors. 
For larger $p$, such as $p=1000$, this advantage is nearly non-existent. On Tiny ImageNet, we see similar performance gains to the small $p$ setting, which can be explained by the fact that, like most real datasets, the data can be well-approximated as lying in a low-dimensional subspace.

\section{PROOF OF THEOREM~\ref{thm:$k$-NN}}

The proof of Theorem \ref{thm:$k$-NN} relies on relating the sample  complexity to the computational complexity. We use that the computational complexity of the adaptive $k$-NN algorithm is no more than $\log (n)$ times the sample complexity of the adaptive $k$-NN algorithm. 
To see this, note that, as discussed in Section~\ref{sec:iteration-complexity}, each iteration has computational cost at most $O(\log (n))$. The initialization of the heaps at the start of the algorithm can be done in $O(n)$ computations using Floyd's algorithm, which is smaller or equal to the sample complexity. 
As a consequence, the computational complexity of the adaptive $k$-NN algorithm is no larger than $O(\log (n))$ times the sample complexity.

For notational convenience, we assume throughout that the distances are ordered so that
\[
d_1 \leq d_2 \leq \ldots \leq d_n.
\]

We begin by showing that the estimate
$\hatdist_i(\NC_i)$ is guaranteed to be $\alpha_i$-close to
$\dist_i$, for all $i$, with high probability.  

\begin{lemma}[{\citealp[Lemma~7]{kaufmann_complexity_2016}}]
\label{lem:probound}
For any $\delta \in (0,0.05)$, with probability at least $1 - \delta$, the event 
\begin{align}
\label{eq:taubounded}
\eventscore \defeq \bigg\{ 
&\left| \hatdist_i(\timeind) - \dist_i \right|
\leq \alpha_i, \; \forall i \in [\n], \; \forall \timeind \geq 1 \bigg\}
\end{align}
occurs. 
The statement continues to hold for any $\delta \in (0,1)$ with 
$\alpha_i = \alpha(T_i) = \sqrt{ \frac{2\beta(T_i,\delta')}{T_i} }$, 
$\beta(\timeind,\delta') = 2 \log(125 \log(1.12 \timeind)/\delta')$. 
\end{lemma}

Lemma~\ref{lem:probound} is a non-asymptotic version of the law
of the iterated logarithm from \citep{kaufmann_complexity_2016} and \citep{jamieson_lil_2014}. 
Note that the lemma uses that $\hatdist_i(t)$ is a sum of $t$ independent random variables, each bounded between $0$ and $1$, and $\EX{\hatdist_i(t)} = \dist_i$. 

We first show that, on the event $\eventscore$ defined in equation~\eqref{eq:taubounded}, the set $\hatS$ contains the $\k$ nearest neighbors. 
On the event $\eventscore$, we have by the termination condition~\eqref{eq:termination} (which is satisfied since the algorithm has terminated) that the items in the set 
$\hatS_\close = \{(1),\ldots (k)\}$ have smaller distances than the items in the set 
$\hatS_\far = \{(k+1+h),\ldots,(n)\}$. 
Because there are $k$ distances that are smaller than the distances in $\hatS_\far$, the set $\hatS_\far$ cannot contain any of the $k$-nearest neighbors, i.e., $\hatS_\far \subset \{1, \ldots, \n\} \setminus \S_\close$. Thus $\S_\close \subset \hatS$.

We next show that on the event $\eventscore$, the adaptive $k$-NN algorithm terminates after the desired number of samples. 
Let $\gamma \defeq \frac{\dist_{\k} + \dist_{\k + 1 + \h}}{2}$, and define the event that item $i$ is bad as 
\[
\bad(i) = 
\begin{cases}
\hatdist_i  > \gamma - 3\alpha_i, & i \in \{1,\ldots, \k \} \\
\hatdist_i  < \gamma + 3\alpha_i, & i \in \{\k + 1 + \h,\ldots, \n \} \\
\alpha_i  > \frac{\dist_{\k+1+h} - \dist_{\k}}{4}, & \text{otherwise}. \\
\end{cases}
\]

\begin{lemma}
\label{lem:oneisbad}
If $\eventscore$ occurs and the termination condition~\eqref{eq:termination} is false, 
then either $\bad(\dind_1)$ or $\bad(\bup_2)$ occurs. 
\end{lemma}

Lemma~\ref{lem:oneisbad} is proved in Section~\ref{sec:oneisbad-proof} in the supplementary material. Given Lemma~\ref{lem:oneisbad}, we can complete the proof in the following way. 
For an item $i$, define 
\[
\Delta_i 
=
\begin{cases}
\dist_{\k+1+\h} - \dist_i, 
& i \in \{1, \ldots, \k \}  \\
\dist_i - \dist_{\k},
&  i \in \{ \k + 1 + \h, \ldots, \n \} \\
\dist_{\k+1+\h} - \dist_{\k},
& \text{otherwise}, \end{cases}
\]
and let $\NCb_{i}$ be the smallest integer $u$ 
satisfying the bound
$
\alpha(u)
\leq \Delta_i/8
$. 
A simple calculation (see Section~\ref{sec:factuneedy} in the supplementary material) yields the following fact.

\begin{fact}
\label{fact:uneedy}
On the event $\eventscore$, if $\NC_i \geq \NCb_{i}$ holds, then $\bad(i)$ does not occur.
\end{fact}

Let $\timeind \geq 1$ be the $\timeind$-th iteration of the steps in the algorithm, and
let $\dind_1$ and $\bup_2$ be the two items selected in the algorithm. Note that in each iteration only those distances are estimated. 
By Lemma~\ref{lem:oneisbad}, we can therefore bound the total number distance estimate updates by 
\begin{align}
&2\sum_{t=1}^{\infty} 
\ind{
\bad(\dind_1)
\cup
\bad(\bup_2)
} \nonumber \\
&\leq
2\sum_{\timeind=1}^{\infty} 
\sum_{i=1}^\n
\ind{
(i = \dind_1 \cup i = \bup_2)
\cap \bad(i)
} \nonumber \\
&\mystackrel{(i)}{\leq}
2
\sum_{\timeind=1}^{\infty}
\sum_{i=1}^\n
\ind{
(i = \dind_1 \cup i = \bup_2)
\cap \NC_i(t) \leq \NCb_{i}
} \nonumber \\
&\mystackrel{(ii)}{\leq}
2
\sum_{i=1}^{\n}
\NCb_{i}.
\label{eq:indtit}
\end{align}
For inequality~(i), we used Fact~\ref{fact:uneedy},
and inequality~(ii) follows because 
$\NC_{i}(t) \leq \NCb_{i}$ can only be true for 
$\NCb_{i}$ iterations $\timeind$. 

We conclude the proof by noting that the definition of $\alpha(\cdot)$ yields the following upper bound (see Section~\ref{sec:bound-T} in the supplementary material).

\begin{fact}
\label{fact:bound-T}
For $c$ sufficiently large,
\begin{align}
\NCb_{i}
&\leq
c
\log\left(\frac{\n}{\delta}\right)
\frac{\log(2\log(2/ \Delta_i  ))}{ \Delta_i^2 }.
\end{align}
\end{fact}

Applying this inequality to the right-hand side of equation~\eqref{eq:indtit} above concludes the proof. 

\section{PROOF OF THEOREM \ref{thm:necessity}}

\newcommand\nuone{\nu}
\newcommand\nutwo{\nu'}
\newcommand\eventone{W}
\newcommand\eventtwo{W'}
\newcommand\bernrv{X}
\newcommand\m{\ell}
\newcommand\setM{\mathcal M}
\newcommand{\sigmaalgebra}{\mc F}
\newcommand\stoptime{\xi} \newcommand\numcmp{N} \newcommand\kl{\mathrm{kl}} \newcommand\KL{\mathrm{KL}} \newcommand{\event}{\ensuremath{\mathcal{E}}}

Consider an algorithm $\alg$ that can only interact with the data by sampling a index $j$ uniformly at random and then obtaining $|[\vx_i]_j - [\vx]_j|$ in response. 
Because the points satisfy $[\vx]_j \in \{-1/2, 1/2\}$,
this is equivalent to drawing from a distribution $\nu_i$ corresponding to a binary $\{0, 1\}$ random variable that has mean $\dist_i$. 
The problem of finding a subset containing the $\k$ nearest neighbors then corresponds to the problem of identifying a subset of all $n$ distribution consisting of $\k+\h$ distributions containing the $\k$ smallest means. Here we will consider only the case where $\h < \k$. This is an \emph{approximate} version of the bottom-$\k$ identification problem in the bandit literature~\citep{kalyanakrishnan_pac_2012}. 
Thus, to prove Theorem \ref{thm:necessity}, we provide a lower bound on the sample complexity required to find a subset of $\k+\h$ distributions containing the $\k$ smallest means, and this lower bound is also a lower bound on the computational complexity.

Towards this goal, we first introduce some notation required
to state a useful lemma~\cite[Lemma~1]{kaufmann_complexity_2016} from
the bandit literature. 
Let $\nu = \{\nu_i\}_{i=1}^n$ be a collection
of $n$ probability distributions, each supported on $\{0, 1\}$.  Consider an algorithm $\alg$, that, at times
$\timeind=1,2,\ldots$, selects the index $i_\timeind \in [n]$ and
receives an independent draw $\bernrv_\timeind$ from the distribution
$\nu_{i_\timeind}$ in response.  Algorithm $\alg$ may select
$i_\timeind$ only based on past observations; that is,
$i_\timeind$ is  \mbox{$\mc F_{\timeind-1}$-measurable}, where
$\sigmaalgebra_\timeind$ is the $\sigma$-algebra generated by
$i_1,\bernrv_{1},\ldots,i_\timeind,\bernrv_{\timeind}$.
Algorithm $\alg$ has a stopping rule $\stoptime$ that determines the
termination of $\alg$.  We assume that $\stoptime$ is a stopping time
measurable with respect to $\sigmaalgebra_\timeind$ and obeying
$\PR{\stoptime < \infty} =1$.

Let $\numcmp_i(\stoptime)$ denote the total number of times index $i$
has been selected by the algorithm $\alg$ (until termination).  For
any pair of distributions $\nu$ and $\nu'$, we let $\KL(\nu,
\nu')$ denote their Kullback-Leibler divergence, and for any $p, q \in
    [0,1]$, let $\kl(p,q) \defeq p \log \frac{p}{q} + (1-p) \log
    \frac{1-p}{1-q}$ denote the Kullback-Leiber divergence between two
    binary random variables with success probabilities $p, q$.

With this notation, the following lemma relates the cumulative number
of comparisons to the uncertainty between the actual distribution
$\nu$ and an alternative distribution $\nu'$.

\begin{lemma}[{\citealp[Lemma~1]{kaufmann_complexity_2016}}]
Let $\nu,\nu'$ be two collections of $n$ 
probability distributions on
$\reals$.  Then for any event $\event \in \sigmaalgebra_\stoptime$ with
$\PR[\nu]{\event} \in (0,1)$, we have
\begin{align}
\sum_{i=1}^n \EX[\nu]{\numcmp_i(\stoptime)} \KL(\nu_i,
\nu_i') \geq \kl(\PR[\nu]{\event}, \PR[\nu']{\event}).
\end{align}
\label{lem:changemeasure}
\end{lemma}

In our setting, since $\nu_i$ and $\nu_i'$ are binary distributions, $\KL(\nu_i, \nu_i') = \kl(\dist_i, \dist_i')$. Let us now use Lemma~\ref{lem:changemeasure} to prove
Theorem~\ref{thm:necessity}. 

Let $\S_1(\nu)$ be the set of $k$ distributions out of the $n$ distributions $\nu = (\nu_1,\ldots,\nu_n)$ with the smallest means. 
Define the event 
\[
\event \defeq \left\{ \S_1(\nu) \subset \hatS \right\},
\]
corresponding to success of the algorithm $\alg$. 
Recalling that
$\stoptime$ is the stopping rule of algorithm $\alg$, we are
guaranteed that $\event \in \sigmaalgebra_\stoptime$. 
Let $\setM \defeq \{ \m_1,\ldots, \m_{\h+1} \}$ be a set of distinct items from the set of $n-k$ distributions with the largest means denoted by $\S_2(\nu)$. 
We next construct an alternative distribution $\nu'$ such that under that distribution, $\m_1,\ldots, \m_{\h+1} \notin \S_2(\nu')$, or equivalently, that $\setM \subseteq \S_1(\nu')$.
Since we assume algorithm $\alg$ succeeds for any distribution with probability at least $1-\delta$, we have both $\PR[\nu]{\event} \geq 1-\delta$ and $\PR[\nu']{\event} \leq \delta$. 
To see this, note that if $\alg$ succeeds under $\nu'$, then $\setM \subset \hatS$. As such, there can be at most $\k - 1$ elements of $\S_1(\nu)$ in $\hatS$, which means that $\event$ does not occur.

It follows that
\begin{align}
&\kl(\PR[\nu]{\event}, \PR[\nu']{\event}) 
\geq \kl(\delta,1-\delta) \nonumber \\
&= (1-2\delta) \log \frac{1-\delta}{\delta} \geq \log \frac{1}{2\delta},
\label{eq:lbdApl2}
\end{align}
where the last inequality holds for $\delta \leq 0.15$. 
It remains to specify the alternative distribution $\nu'$. The alternative distribution is defined as
\begin{align*}
\nu_i'
&= 
\begin{cases}
\nu_{\k-\h}, & i \in \setM \\
\nu_i, & \text{otherwise}.
\end{cases}
\end{align*}
To be most precise on avoiding ties, for $\m \in \setM$, one should take $\nu_\m' = \nu_{\k-\h} - \varepsilon$ for some $\varepsilon > 0$ and let $\varepsilon\to 0$, but we omit carrying out the associated technical details in this proof. 
It follows that, under the distribution $\nu'$, the items in the set $\setM$ are not among the items with the $n-k$ largest means which ensures that $\setM \cap \S_2(\nu') = \emptyset$. 

Let $\numcmp_\m$ be the total number of draws from the distribution $\nu_\m$. 
We have that
\begin{align}
\sum_{\m \in \setM}
\kl(\nu_{\m}, \nu_{\m}') \EX[\nu]{\numcmp_\m} 
& \mystackrel{(i)}{=} \sum_{i=1}^n 
\EX[\nu]{\numcmp_{i}} \kl(\nu_{i}, \nu_{i}') \nonumber \\
& \mystackrel{(ii)}{\geq} \kl(\PR[\nu]{\event}, \PR[\nu']{\event})
\nonumber \\
\label{eq:bylbdApl2}
& \geq \log \frac{1}{2\delta}.
\end{align}
Here step~(i) follows from the fact that $\kl(\nu_{i},
\nu_{i}')=0$ for all $i \notin [n] \setminus \setM$ by definition of the $\nu'_{i}$, and step~(ii) follows from Lemma~\ref{lem:changemeasure}. Finally, inequality~\eqref{eq:bylbdApl2} follows from inequality~\eqref{eq:lbdApl2}.

We next upper bound the KL divergence on the left hand side of
inequality~\eqref{eq:bylbdApl2}.
Using the inequality $\log x \leq
x-1$, valid for $x>0$, we have that for $\m \in \setM$,
\begin{align}
\kl(\nu_{\m}, \nu_{\m}')
&\leq 
     \kl_{\m},
  \quad \kl_{\m}
  \defeq
  \frac{(\dist_{\m} - \dist_{\k - \h})^2}{\dist_{\k-\h}
  (1-\dist_{\k-\h}) }.
\label{eq:klbscore2}
\end{align}

Applying inequality~\eqref{eq:klbscore2} to the left hand side of inequality~\eqref{eq:bylbdApl2} yields 
\begin{align}
\label{eq:constraints}
\sum_{\m \in \setM} \kl_\m  \EX[\nu]{\numcmp_\m}  
\geq 
\log \frac{1}{2\delta},
\end{align}
which is valid for each subset $\setM \subseteq \S_2(\nu)$ of cardinality $\h+1$.

We can therefore obtain a lower bound on 
$\sum_{i \in \S_2(\nu)} \EX[\nu]{\numcmp_i}$ 
by solving the minimization problem:
\begin{align}
&\underset{e_\m \geq 0}{\text{minimize}} \sum_{\m \in \S_2(\nu)} e_\m 
\quad \text{subject to} \;
\sum_{\m \in \setM} \kl_\m e_\m
\geq \log \frac{1}{2\delta} \nonumber \\
&\text{  for all $\setM \subseteq \S_2(\nu)$ of cardinality $\h+1$}.
\end{align}
Since the $\kl_\m$ are increasing in $\m$ (recall that we assume the distances to be ordered such that $d_1 \leq d_2 \leq \ldots \leq d_\n$) the solution to this optimization problem is $e_{\k+1},\ldots,e_{\k+\h} = 0$ and 
$e_{\m} = \log(1/2\delta)/\kl_\m$ for $\m \geq \k + 1 + \h$. 

Using an analogous line of arguments for 
items in the set $\setS_1(\nu)$ 
(see Section~\ref{sec:second-alternate-distribution} in the supplemental material), we arrive at the lower bound
\begin{align*}
\log \frac{1}{2\delta}
\sum_{i = 1}^{k-\h}
\frac{d_{\k + 1 + \h} (1-d_{\k + 1 + \h}) }{( d_i  - d_{k+1+\h} )^2} \\
+ 
\log \frac{1}{2\delta} \sum_{i = k+1+\h}^{n}
\frac{d_{\k - \h} (1 - d_{\k - \h}) }{( \d_{k-\h}  - d_{i} )^2}
\end{align*}
on the number of comparisons. 
This concludes the proof.

\subsubsection*{Acknowledgements}
We thank the anonymous reviewers for their constructive feedback. DL and RB are partially supported by NSF grants IIS-17-30574 and IIS-18-38177, AFOSR grant FA9550-18-1-0478, ARO grant W911NF-15-1-0316, ONR grants N00014-17-1-2551 and N00014-18-12571, DARPA grant G001534-7500, and DOD Vannevar Bush Faculty Fellowship (NSSEFF) grant N00014-18-1-2047. RH is partially supported by NSF award IIS-1816986.

\bibliographystyle{abbrvnatnourl}
\balance

\clearpage
\appendix

\begingroup
\allowdisplaybreaks

\section{PROOFS OF LEMMAS AND FACTS}
\subsection{Proof of Lemma \ref{lem:oneisbad}}
\label{sec:oneisbad-proof}

\newcommand\good{\mc{E}_\mathrm{good}}
The proof is very similar to the proof of Lemma 2 of \cite{heckel_approximate_2018}. There are several cases of $\dind_1$ and $\bup_2$ to consider.
We will show each by contradiction, starting with the assumption that the termination condition is false and both $\bad(\dind_1)$ and $\bad(\bup_2)$ do not occur, all under the event $\eventscore$. Let $\good(i)$ denote the complement of $\bad(i)$. It also will be useful to define the quantity
\begin{align}
    m_2 = \argmax_{i \in \{(\k + 1), \ldots, (\k + \h)\}} \alpha_i
\end{align}
such that $\bup_2 = \argmax_{i \in \{m_2, \dind_2\}} \alpha_i$.

\begin{enumerate}
    \item When $\dind_1 \leq \k$ and $\bup_2 > \k + \h$, we have by $\good(\dind_1)$ that
    \begin{align}
        \hatdist_{\dind_1} + \alpha_{\dind_1} < \hatdist_{\dind_1} + 3 \alpha_{\dind_1} \leq \gamma
    \end{align}
    and similarly that ${\hatdist_{\bup_2} - \alpha_{\bup_2} > \gamma}$ by $\good(\bup_2)$. Since $\hatdist_{\dind_2} - \alpha_{\dind_2} \geq \hatdist_{m_2} - \alpha_{m_2}$, we have that $\hatdist_{\dind_2} - \alpha_{\dind_2} > \gamma$ in both the case that $\bup_2 = m_2$ and $\bup_2 = \dind_2$. Together, this implies that the termination condition \eqref{eq:termination} is true, which violates our assumption.

    \item 
    \label{case:q1-far-b2-mid}
    When $\dind_1 \leq \k$ and $\k < \bup_2 \leq k + h$, we have first by $\good(\dind_1)$ that ${\hatdist_{\dind_1} + 3\alpha_{\dind_1} \leq \gamma}$. Starting from here, and using the definition of $\dind_1$, we have for all $i \in \hatS_\close$,
\begin{align}
    \gamma &\geq \hatdist_{\dind_1} + \alpha_{\dind_1} + 2\alpha_{\dind_1}\nonumber \\
    &\geq \hatdist_i + \alpha_i + 2\alpha_{\dind_1}\nonumber \\
    &\geq \dist_i + 2\alpha_{\dind_1}\nonumber \\
    &> \dist_i.
    \label{eq:lem2-bound-close}
\end{align}
Now we let $\Delta$ denote $\dist_{\k+1+\h} - \dist_\k$. By definition of $\bup_2$, using $\good(\bup_2)$, we have that ${\alpha_j \leq \Delta / 4}$ for all ${j \in \hatS_\middle \cup \{\dind_2\}}$. Then we can start from ${\gamma > \hatdist_{\dind_1} + \alpha_{\dind_1}}$ to conclude that for all ${j \in \hatS_\middle \cup \{\dind_2\}}$,
\begin{align}
    \gamma &> \hatdist_{\dind_1} + \alpha_{\dind_1} \nonumber \\
    &\overset{(i)}{>} \hatdist_{\dind_2} - \alpha_{\dind_2} \nonumber \\
    &\geq \hatdist_{\dind_2} - \frac{\Delta}{4} \nonumber \\
    &\geq \hatdist_j - \frac{\Delta}{4} \nonumber \\
    &\geq \dist_j - \alpha_j - \frac{\Delta}{4} \nonumber \\
    &\geq \dist_j - \frac{\Delta}{2},
    \label{eq:lem2-bound-mid}
\end{align}
where $(i)$ comes from our assumption that the terminating condition \eqref{eq:termination} is false.
Combining \eqref{eq:lem2-bound-close} and \eqref{eq:lem2-bound-mid} along with ${\gamma + \Delta/2 = d_{\k+1+\h}}$, we obtain that ${\dist_{\k+1+\h} > d_i}$ for all $i \in \hatS \cup \{\dind_2\}$, which is a contradiction, since there can be at most $\k + \h$ values of $\dist_i$ that are smaller than $\dist_{\k+1+\h}$.

\item When $\k < \dind_1 \leq \k + \h$ and $\bup_2 > \k + \h$, the case is similar to the previous case, except that we need to bound $\alpha_i$ for $i \in \hatS_\middle$ in a different way. By $\good(\bup_2)$, $\hatdist_{\dind_2} \geq \hatdist_{\bup_2}$, and $\alpha_{\bup_2} \geq \alpha_{\dind_2}$, we have analogously to \eqref{eq:lem2-bound-close}, for all $i \in \hatS_\far$,
\begin{align}
    \gamma &\leq \hatdist_{\bup_2} - 3 \alpha_{\bup_2} \nonumber \\
    &\leq \hatdist_{\dind_2} - \alpha_{\dind_2} - 2\alpha_{\bup_2} \nonumber \\
    &\leq \dist_i - 2 \alpha_{\bup_2}.
\end{align}
Equivalently, $\dist_i \geq \gamma + 2\alpha_{\bup_2}$. Since there are ${n - \k - \h}$ values of $i$ for which this inequality holds, it must hold for $\dist_{\k+1+\h}$, so we obtain
\begin{align}
    \alpha_{\bup_2} \leq \frac{\dist_{\k+1+\h} - \gamma}{2} 
    = \frac{\Delta}{4}.
\end{align}
By definition of $\bup_2$, $\alpha_i \leq \Delta/4$ for all $i \in \hatS_\middle \cup \{\dind_2\}$, and a contradiction can be reached similarly as in case \ref{case:q1-far-b2-mid}.

\item For the case when both $\dind_1, \bup_2 \in \{\k+1, \ldots, \k + \h\}$, we first show that at least one of ${\gamma < \hatdist_{\dind_1} + \alpha_{\dind_1}}$ or $\gamma > \hatdist_{\dind_2} - \alpha_{\dind_2}$ is true. 
To see this, first suppose the former is false. Then using that the terminating condition \eqref{eq:termination} is false, we have
\begin{align}
    \gamma \geq \hatdist_{\dind_1} + \alpha_{\dind_1} > \hatdist_{\dind_2} - \alpha_{\dind_2}.
\end{align}
Now that we know that at least one of these inequalities holds, and we proceed similarly for each. 
First suppose that the former inequality, ${\gamma < \hatdist_{\dind_1} + \alpha_{\dind_1}}$, holds. 
Using that by $\good(\dind_1)$ and $\good(\bup_2)$ we have $\alpha_i \leq \Delta/4$ 
for all $i \in \{\dind_1, \dind_2\} \cup \hatS_\middle$, we have that, for all $i \in \{\dind_1, \dind_2\} \cup \hatS_\middle$,
\begin{align}
    \label{eq:lem3-case4-mid}
    \gamma &< \hatdist_{\dind_1} + \alpha_{\dind_1} \nonumber \\
    &\leq \hatdist_i + \alpha_{\dind_1} \nonumber \\
    &\leq d_i + \alpha_i + \alpha_{\dind_1} \nonumber \\
    &\leq d_i + \frac{\Delta}{2}.
\end{align}
We also have for all $j \in \hatS_\far$ that
\begin{align}
    \label{eq:lem3-case4-far}
    \gamma &< \hatdist_{\dind_1} + \alpha_{\dind_1} \nonumber \\
    &\leq \hatdist_{\dind_2} - \alpha_{\dind_2} + \alpha_{\dind_2} + \alpha_{\dind_1} \nonumber \\
    &\leq \hatdist_j - \alpha_j + \alpha_{\dind_2} + \alpha_{\dind_1} \nonumber \\
    &\leq \dist_j + \alpha_{\dind_2} + \alpha_{\dind_1} \nonumber \\
    &\leq \dist_j + \frac{\Delta}{2}.
\end{align}
Combining \eqref{eq:lem3-case4-mid} and \eqref{eq:lem3-case4-far}, we have that $d_i > d_{\k}$ for all $i \in \{\dind_1\} \cup \hatS_\middle \cup \hatS_\far$, which is a contradiction, since at most $n - \k$ values of $i$ can satisfy this inequality.

The case that $\gamma > \hatdist_{\dind_2} - \alpha_{\dind_2}$ is entirely analogous.

\item When $\dind_1 > \k + \h$ or $\bup_2 \leq k$, we can make similar arguments to the previous cases to reach a contradiction.

\end{enumerate}

\subsection{Proof of Fact \ref{fact:uneedy}}
\label{sec:factuneedy}
First, when $i \leq k$, we have
\begin{align}
    \hatdist_i + 3\alpha_i 
    &\leq \dist_i + 4 \alpha_i \nonumber \\
    &\leq \dist_i + \frac{\Delta_i}{2} \nonumber \\
    &\leq \frac{\dist_{\k+1+\h} + \dist_i}{2} \leq \gamma,
\end{align}
where the last inequality uses $\dist_i \leq \dist_\k$, so $\bad(i)$ does not occur. This is similarly shown for $i > \k + \h$. For ${\k < i \leq \k + \h}$, that $\bad(i)$ does not occur follows immediately from $\alpha_i \leq \Delta_i/8 \leq \Delta_i/4$.

\subsection{Proof of Fact~\ref{fact:bound-T}}
\label{sec:bound-T}
Recalling that $\alpha_i(u) = \sqrt{ \frac{2\beta(u,\delta/\n) }{ u} }$, 
 at ${\alpha_i(u) = \Delta_i/8}$ we have that $u = 2(\Delta_i/8)^{-2} \beta(u, \delta')$, so we need to bound the greatest fixed point $u^*$ of 
 \begin{align*}
 f(u) = 2(\Delta_i/8)^{-2} \beta(u, \delta').
 \end{align*}
 Let $u_0 = 2(\Delta_i/8)^{-2}$, and note that for all $u \geq u_0$,
  \begin{align}
     f'(u)
     &= \frac{2(\Delta_i/8)^{-2}(2)}{u \log(1.12 u)} \nonumber \\
          &\leq \frac{2(\Delta_i/8)^{-2}(2)}{2(\Delta_i/8)^{-2} \log((1.12) 2(\Delta_i/8)^{-2})} \nonumber \\
          &\leq \frac{2}{\log ((1.12) 32)} \nonumber \\
          &< 1.
 \end{align}
The second inequality holds because $\Delta_i \leq 2$. Suppose that $u^* > u_0$. Using Taylor's theorem, we have that for some $z \geq u_0$,
\begin{align}
    f(u_0) &= f(u^*) + f'(z) (u_0 - u^*) \nonumber \\
    &= u^*(1 - f'(z)) + u_0 f'(z).
\end{align}
Then 
\begin{align}
    u^* &= \frac{f(u_0) - u_0 f'(z)}{1 - f'(z)} \nonumber \\
    &\leq \frac{f(u_0)}{1 - f'(u_0)}.
\end{align}
So, we can bound the greatest fixed point of $f$ as
\begin{align}
    u^* &\leq \max \left\{ u_0, \frac{f(u_0)}{1 - f'(u_0)} \right\} \nonumber \\
    &= 2(\Delta_i/8)^{-2} \max \left\{ 1, \frac{\beta(2(\Delta_i/8)^{-2}, \delta')}{1 - 2/\log ((1.12)32)} \right\} \nonumber \\
        &= c_1 \Delta_i^{-2} \beta(2(\Delta_i/8)^{-2}, \delta'),
    \end{align}
where $c_1 = 128 / (1 - 2/\log ((1.12)32))$.
Since $\NCb_i \leq u^* + 1$, letting $c_2 = c_1 + 1$,
\begin{align}
    \NCb_{i} \leq \frac{c_2}{\Delta_i^2} \log \left(125\frac{n}{\delta} \log \left( \frac{(1.12)128}{\Delta_i^2} \right) \right).
    \end{align}
Then for $c$ sufficiently large,
\begin{align}
\NCb_{i}
&\leq
c
\log\left(\frac{\n}{\delta}\right)
\frac{\log(2\log(2/ \Delta_i  ))}{ \Delta_i^2 }.
\end{align}

\section{ADDITIONAL THEOREM \ref{thm:necessity} PROOF DETAILS}

\label{sec:second-alternate-distribution}

In this section we provide details on bounding  $\sum_{i \in \setS_1(\nu)} \EX[\nu]{\numcmp_i}$
that we omitted in the proof of Theorem~\ref{thm:necessity}. 
We consider the set $\setM = \{\m_1, \ldots, \m_{\h+1}\} \subseteq \setS_1(\nu)$ and construct an alternative distribution $\nu'$ such that under that distribution $\setM \subseteq \setS_2(\nu')$. Then under $\nu'$, if $\alg$ succeeds, then at most $\h$ elements of $\setS_2(\nu')$ can be in $\hatS$, meaning that at least one element of $\setM$ is not in $\hatS$ and that $\event$ does not occur. So, if $\alg$ succeeds with probability at least $1 - \delta$, then both $\PR[\nu]{\event} \geq 1 - \delta$ and $\PR[\nu']{\event} \leq \delta$.

Our alternative distribution $\nu'$ is defined as
\begin{align*}
    \nu_i' = \begin{cases}
    \nu_{\k + 1 + \h}, & i \in \setM \\
    \nu_i, & \text{otherwise}.
    \end{cases}
\end{align*}
Again, to avoid ties, for $\m \in \setM$, one should take $\nu_\m' = \nu_{\k + 1 
+\h} + \varepsilon$ and let $\varepsilon \to 0$, but we omit this detail. The remainder of the arguments are entirely analogous to the case shown previously, giving us the bound
\begin{align*}
    \sum_{i \in \setS_1(\nu)} \EX[\nu]{\numcmp_i} \geq 
    \log \frac{1}{2\delta}
\sum_{i = 1}^{k-\h}
\frac{d_{\k + 1 + \h} (1-d_{\k + 1 + \h}) }{( d_i  - d_{k+1+\h} )^2}.
\end{align*}

\endgroup

\balance

\end{document}